\documentclass[12pt]{article}

\usepackage{amsmath,amssymb,amsthm}
\usepackage[mathscr]{eucal}

\usepackage{graphicx}

\DeclareMathOperator{\Tr}{Tr}

\DeclareMathOperator{\Si}{Si}
\DeclareMathOperator{\Ci}{\tilde{Ci}}

\newcommand{\pl}{\partial}
\newcommand{\ol}{\overline}

\newcommand{\DD}{\mathcal{D}}
\newcommand{\HH}{\mathscr{H}}
\newcommand{\II}{\mathrm{I}}

\newcommand{\IM}{\mathcal{I}m}
\newcommand{\RR}{\mathbb{R}}
\newcommand{\CC}{\mathbb{C}}

\newcommand{\OO}{\mathcal{O}}

\newcommand{\lL}{\mathfrak{l}}
\newcommand{\JJ}{\mathrm{J}}
\newcommand{\sigmab}{\sigma}
\newcommand{\Hilbert}{\mathrm{H}}
\newcommand{\HSop}{\mathscr{E}}

\newcommand{\printbookitem}[5]{#1, \textit{#3}, #4, #2, #5.}
\newcommand{\printpreprintitem}[4]{#1, ``#3,'' #4 (#2).}
\newcommand{\printjournalitem}[7][]{#2, ``#4,'' #5 \textbf{#6} #1 (#3), #7.}

\renewcommand{\vec}[1]{\boldsymbol{#1}}

\begin{document}

\thispagestyle{empty}
\begin{center}
 { \Large \textbf{
Some solutions to the eigenstate equation for the free quantum field
Hamiltonian in the Schr\"odinger representation
}}\\
\vspace{0.3cm}
T.~A.~Bolokhov\\
\textit{St.\,Petersburg Department of V.\,A.\,Steklov Mathematical Institute}\\
\textit{         Russian Academy of Sciences,}\\
\textit{27 Fontanka, St.\,Petersburg, Russia 191023}\\
email: timur@pdmi.ras.ru
\end{center}

\begin{abstract}
    Using closed positive extensions of the quadratic form
    in the potential term we provide alternative solutions
    to the eigenstate equation for the free
    quantum field Hamiltonian in the Schr\"o\-din\-ger representation.
    We show that admissible extensions stem from the
    singular behaviour of the quantum field simultaneously in at least
    two points,
    the distance between the latter being limited by the extension parameter.
    Including position of singularities into dynamical model
    we provide example of quantum field
    evolved by the action of the free Hamiltonian
    interacting with external sources via functional boundary conditions.
\end{abstract}

\medskip


\newpage
\tableofcontents

\section{Introduction}
    The Hamilton operator of the free scalar quantum field in the Schr\"odinger
    representation is the sum of the second variational derivative
    and the quadratic form of the Laplace operator
\cite{Hatfield}
\begin{equation}
\label{H0}
    \HH  = \int_{\RR^{3}}d^{3}x \, \bigl(
    -\frac{\delta}{\delta A(\vec{x})} \frac{\delta}{\delta A(\vec{x})}
     + (\pl_{m} A)^{2}
    \bigr) , \quad A(\vec{x}) = \ol{A(\vec{x})}.
\end{equation}
    This operator acts on the set of twice differentiable functionals
    defined on the domain of the potential term.
    In order to construct the ground state of
$ \HH $
    one needs to employ spectral decomposition of the quadratic form
    by fixing the scalar product
\begin{equation}
\label{prodx}
    (f(\vec{x}), g(\vec{x})) = \int_{\RR^{3}} d^{3}x \, \ol{f(\vec{x})}
	g(\vec{x}),
\end{equation}
    and expressing the form in the terms of the corresponding
    self-adjoint operator
\begin{equation*}
    \lL[A] = \int_{\RR^{3}} (\pl_{m}A(\vec{x}))^{2} \, d^{3}x 
	= \ol{(A, LA)}, \quad L = -\frac{\pl^{2}}{\pl x_{k}^{2}}
\end{equation*}
    (the bar over quadratic form denotes its closure
    \textit{w.\,r.}~to the scalar product
(\ref{prodx})).
    Then the functional of the ground state of the operator
(\ref{H0})
    is given by the Gaussian of the quadratic form
$ \omega[A] $
    corresponding to the square root of
$ L $
\begin{equation}
\label{Omega0}
    \Omega_{0}(A) = \exp\{-\frac{1}{2}\omega[A]\} ,\quad
	\omega[A] = \ol{(A,L^{1/2}A)}.
\end{equation}
    Due to the equation
\begin{equation*}
    \frac{1}{4}\int_{\RR^{3}}d^{3}x \, \bigl(\frac{\delta}{\delta A(\vec{x})}
	\omega[A] \bigr)^{2} = (L^{1/2}A, L^{1/2}A) = \lL[A],
    \quad A \in \DD(\lL) = \DD(L^{1/2}),
\end{equation*}
    the action of
$ \HH $
    on the functional
$ \Omega_{0}(A) $
    is reduced to multiplication by the scalar
$ \Tr L^{1/2} $
\begin{multline*}
    \HH \, \Omega_{0}(A) = \int_{\RR^{3}}d^{3}x \, \bigl(
    -\frac{\delta^{2}}{\delta A(\vec{x})^{2}} + (\pl_{m} A)^{2}
    \bigr) \, \exp\{-\frac{1}{2}\omega[A]\} =\\
    = \frac{1}{2}
    \int_{\RR^{3}} d^{3}x\,\bigl(\frac{\delta^{2}}{\delta A(\vec{x})^{2}}
	\omega[A] \bigr) \, \exp\{-\frac{1}{2}\omega[A]\}
    = (\Tr L^{1/2}) \, \Omega_{0}(A) .
\end{multline*}
    One may note that the flat integral in the first term of
(\ref{H0})
    leads to the flat scalar product in the quadratic form
$ \omega[A] $
    and in this way there is no other option in the definition of
(\ref{prodx}).

    The Hamiltonian
$ \HH $
    has singularities
    in the region where its potential term is infinite, that is
    outside of the domain of the quadratic form 
$ \DD(\lL) $.
    There is no natural measure on the space of functions to describe
$ \DD(\lL) $
    as an open set which does not include boundary, but one may say
    that functions with behaviour
\begin{equation}
\label{ABound}
    A(\vec{x}) =
	\frac{C_{n}}{|\vec{x} - \vec{x}_{n}|^{1/2}} ,
    \quad \vec{x}\to\vec{x}_{n},
    \quad n = 1,\ldots N
\end{equation}
    reside very close to
$ \DD(\lL) $.
    Indeed, for any
$ \epsilon > 0 $,
    if
$ A(\vec{x}) $
    expands around
$ \vec{x}_{n} $
    as
\begin{equation*}
    A(\vec{x}) = \frac{C_{n}}{|\vec{x}-\vec{x}_{n}|^{1/2-\epsilon}},
    \quad \vec{x} \to \vec{x}_{n},
\end{equation*}
    then the integrand in
$ \lL[A] $
    has singularity of the order
\begin{equation*}
    (\pl_{m} A(\vec{x}))^{2} = 
	\frac{C_{n}^{2}}{|\vec{x}-\vec{x}_{n}|^{3-2\epsilon}},
    \quad \vec{x} \to \vec{x}_{n},
\end{equation*}
    and so is locally integrable in
$ \RR^{3} $.
    Meanwhile in the case
$ \epsilon = 0 $
    the integral in the quadratic form diverges logarithmically 
    and thus the Hamiltonian
$ \HH $
    acquires singularity in the vicinity of functions with behaviour
(\ref{ABound}).
    As long as the domain of
$ \omega $
    is wider than 
$ \DD(\lL) $,
    the functional
$ \Omega_{0}(A) $
    is well defined
    in the vicinity of
(\ref{ABound})
    and satisfies specific boundary condition.

    Now assume that
    there exists positive extension
$ \lL_{\kappa}[A] $
    of the quadratic form
$ \lL[A] $
    to some function of behaviour
(\ref{ABound})
\begin{align}
\label{lext1}
    \lL_{\kappa}[A] & = \lL[A] , \quad A \in \DD(\lL),\\
\label{lext2}
    \lL_{\kappa}[B] & < \infty,
	\quad B \stackrel{\vec{x}\to\vec{x}_{n}}{=}
	\frac{C_{n}}{|\vec{x}-\vec{x}_{n}|^{1/2}} .
\end{align}
    Provided that
$ \lL_{\kappa} $
    is closed \textit{w.\,r.}~to the scalar product
(\ref{prodx})
    one may 
    construct corresponding
    self-adjoint operator
$ L_{\kappa} $
    and define its spectral decomposition
    (on extensions of closed semi-bounded quadratic forms
see~\cite{Koshmanenko1}).
    Then one may substitute 
$ \omega[A] $
    in
(\ref{Omega0})
    with the quadratic form of the square root of
$ L_{\kappa} $
    and observe that the Gaussian
\begin{equation}
\label{OmegaK}
    \Omega_{\kappa}(A) = \exp\{-\frac{1}{2}\omega_{\kappa}[A]\}, \quad
	\omega_{\kappa}[A] = \ol{(A,L_{\kappa}^{1/2}A)}
\end{equation}
    also satisfies the eigenstate equation for
$ \HH $
\begin{equation}
\label{HOk}
    \HH \, \Omega_{\kappa}(A)
    = \frac{1}{2}
    \int_{\RR^{3}} d^{3}x\,\bigl(\frac{\delta^{2}}{\delta A(\vec{x})^{2}}
	\omega_{\kappa}[A] \bigr) \, \exp\{-\frac{1}{2}\omega_{\kappa}[A]\} 
    = (\Tr L_{\kappa}^{1/2}) \, \Omega_{\kappa}(A) ,
\end{equation}
    provided that the kernel of
$ \omega_{\kappa} $
    is symmetric and
    the following equation holds
\begin{equation}
\label{lqeq}
    \frac{1}{4}\int_{\RR^{3}}d^{3}x \, \bigl(\frac{\delta}{\delta A(\vec{x})}
	\omega_{\kappa}[A] \bigr)^{2} = (L_{\kappa}^{1/2}A, L_{\kappa}^{1/2}A)
    = \lL_{\kappa}[A] = \lL[A],
    \quad A \in \DD(\lL).
\end{equation}

    In order to verify equations
(\ref{HOk})
    and
(\ref{lqeq})
    one needs to fix the set \textit{w.\,r.}~to which the variational
    derivative is operating.
    The form
$ \omega[A] $
    in
(\ref{Omega0})
    is defined as Fourier preimage of the quadratic form of multiplication
    operator
    and can be reconstructed
    by variation \textit{w.\,r.}~to functions from the domain
    of the quadratic form
$ \lL[A] $.
    This trivial proposition can be written as
\begin{equation*}
    \overline{\omega|_{\DD(\lL)}} = \omega ,
\end{equation*}
    where LHS is the closure of restriction of the form
$ \omega[A] $
    to the set
$ \DD(\lL) $.
    That is for any
$ A\in \DD(\omega) $
    there exists a sequence
$ A_{n}\in \DD(\lL) $
    such that
\begin{equation*}
    || A_{n} -A||_{L_{2}(\RR^{3})} \stackrel{n\to\infty}{\to} 0,\quad
    \omega[A_{n} - A] \stackrel{n\to\infty}{\to} 0.
\end{equation*}
    Thus in order to preserve the action of
$ \HH $
    the same set
$ \DD(\lL) $
    should be used for variations in the eigenstate equation
(\ref{HOk}).
    That is we need to verify that the above closure also holds
    for the quadratic form 
$ \omega_{\kappa}[A] $
\begin{equation}
\label{wkappa}
    \overline{\omega_{\kappa}|_{\DD(\lL)}} = \omega_{\kappa} .
\end{equation}
    As we will see in what follows, proof of this equation requires
    some effort and is not possible for every closable extension
$ \lL_{\kappa} $
    of the quadratic form of the Laplace operator.

    Another essential requirement of the presented construction
    is the positivity of the extension
$ \lL_{\kappa} $.
    Otherwise the operator
$ L_{\kappa} $
    possesses negative eigenspace and the quadratic form
$ \omega_{\kappa} $
    goes imaginary thereon.
    This in turn contradicts the self-adjointness of the action of
$ \HH $
    and does not allow to introduce a scalar product on the set of functionals
    with finite normalization for
$ \Omega_{\kappa} $.

    The construction of solutions to the eigenstate equation
    of the free quantum field by means of the extensions
    of quadratic form in the potential term was proposed in
\cite{TB}.
    There extensions generated by 3-dimensional
$ \delta $-potential
    were used,
    that is extensions to functions with the behaviour
\begin{equation*}
    A(\vec{x}) =
	\frac{C_{n}}{|\vec{x} - \vec{x}_{n}|} ,
    \quad \vec{x}\to\vec{x_{n}}.
\end{equation*}
    However no proof of the possibility to reconstruct the quadratic form of
    the square root by its action on
$ \DD(\lL) $
    for that case is still not provided.

    In this work we develop the scheme of construction of eigenfunctionals
    of the free Hamiltonian by extending the quadratic form of Laplacian to
    functions with behaviour
(\ref{ABound}).
    We will show that closed positive extensions require this behaviour
    in at least 2 points and will derive the
    admissible distance between the latter as a function of the extension
    parameter.
    Then we will verify that extensions in question satisfy equation
(\ref{wkappa}),
    discuss normalization of
$ \Omega_{\kappa} $,
    its Hilbert space
    and possibility to introduce eigenvalues in
(\ref{HOk})
    with continuous dependence on the position of singularities.
    This way the model can be interpreted as dynamical system which includes
    quantum field evolved by the action of the free Hamiltonian and
    external sources -- subject to further quantization.
    We expect that presented construction can be employed in description
    of renormalized asymptotically free quatum field and will formulate
    two propositions in support of this statement.


\section{Resolvents of positive extensions}
\subsection{Extensions of the quadratic form and singular potentials}
    Let us turn to detailed description of extensions of the quadratic form
    of the scalar Laplace operator outlined in
(\ref{lext1}), (\ref{lext2}).
    We will use the Fourier hat to denote the momentum representation
\begin{equation*}
    \hat{A}(\vec{p}) = \frac{1}{(2\pi)^{3/2}} \int_{\RR^{3}} A(\vec{x})
    e^{-i\vec{p}\cdot{\vec{x}}} \, d^{3}x,
    \quad \ol{\hat{A}(\vec{p})} = \hat{A}(-\vec{p})
\end{equation*}
    and leave the quantities without hat for coordinate functions
    or for invariant expressions.
    In this way operator
$ \HH $
    acts in the terms of Fourier images in the following way
\cite{Hatfield},
\cite{Jackiw}
\begin{equation}
\label{HF}
    \HH \, \Omega(A) = \int_{\RR^{3}}d^{3}p \, \bigl(
	-\frac{\delta}{\delta \hat{A}(-\vec{p})}
	    \frac{\delta}{\delta \hat{A}(\vec{p})}
    + p^{2}|\hat{A}(\vec{p})|^{2} \bigr)
    \, \Omega(A).
\end{equation}
    Quadratic forms
$ \lL[A] $
    and
$ \omega[A] $
    are diagonal 
\begin{align*}
    \lL[A] &= \int_{\RR^{3}} p^{2} |\hat{A}(\vec{p})|^{2} \, d^{3}p,
	& L: & \,\hat{A}(\vec{p}) \to p^{2} \hat{A}(\vec{p}), \\
    \omega[A] &= \int_{\RR^{3}} p |\hat{A}(\vec{p})|^{2} \, d^{3}p,
	& L^{1/2}: & \,\hat{A}(\vec{p}) \to p \hat{A}(\vec{p}),
\end{align*}
    while the scalar product connecting forms with corresponding
    operators is still flat
\begin{equation}
\label{prodp}
    (\hat{g}(\vec{p}), \hat{f}(\vec{p})) = \int_{\RR^{3}} 
	\ol{\hat{g}(\vec{p})} \hat{f}(\vec{p}) \, d^{3}p.
\end{equation}
    The diagonal representation of
$ \lL[A] $
    allows to interpret expression
(\ref{HF})
    as a sum of Hamiltonians of quantum harmonic oscillators.
    When each oscillator
    or group of oscillators with nearby frequencies
    is treated separately, it always falls in the domain of
$ \lL[A] $
    and thus has finite potential energy.
    But after multiplication of eigenstates we come to the functional
    of the ground state of the free scalar field
\begin{equation*}
    \Omega_{0}(A) = \exp\{-\frac{1}{2}\omega[A]\}
	= \exp\{-\frac{1}{2}\int_{\RR^{3}} p |\hat{A}(\vec{p})|^{2}\,
	    d^{3}p\},
\end{equation*}
    which has different behaviour.
    The feature of this state is that it
    gives nonzero expectation values for the fields with infinite
    potential energy, by the reason that domain of the potential term
    is smaller than that of
$ \omega $
    (in the UV-region)
\begin{equation*}
    \DD(\lL) \subset \DD(\omega), \quad \DD(\lL) \neq \DD(\omega) .
\end{equation*}
    And the latter hints that there may exist other (non-free) Gaussian
    solutions to
(\ref{HF})
    with quadratic forms defined on other domains non-comparable to
$ \DD(\omega) $
    but including
$ \DD(\lL) $.
    To study these we are going to extend first the quadratic form
    in the potential term of
$ \HH $
    with a view to extract the square root of the corresponding operator.

    Key objects for construction of closable extensions of the quadratic form
$ \lL[A] $
    are the resolvent of the self-adjoint operator
$ L $
\begin{equation*}
    R_{\mu} = (L-\mu \II)^{-1}:
	\ \hat{A}(\vec{p}) \to \frac{1}{p^{2}-\mu} \hat{A}(\vec{p}),
	\quad \mu\in \CC\setminus \RR^{+}\cup 0
\end{equation*}
    and the singular potential
$ \hat{v}(\vec{p}) $
    -- some possibly generalized function which satisfies conditions
\begin{align}
\label{vcond1}
    (R_{\mu}v, R_{\nu}v) &< \infty \\
\label{vcond2}
    (v, R(\mu)v) &
	= \int_{\RR^{3}} \frac{|\hat{v}(\vec{p})|^{2}}{p^{2}-\mu} d^{3}p
	= \infty.
\end{align}
    As candidates for
$ \hat{v}(\vec{p}) $
    we will consider powers of the modulus of
$ \vec{p} $
\begin{equation*}
    \hat{v}_{\alpha}(\vec{p}) = p^{\alpha}.
\end{equation*}
    Such potentials have distinctive singularities
    in the coordinate representation
\begin{equation*}
    v_{\alpha}(\vec{x}) =
    \frac{1}{(2\pi)^{3/2}}
	\int_{\RR^{3}}\hat{v}_{\alpha}(\vec{p}) e^{i\vec{p}\cdot\vec{x}} \,d^{3}p =
    \begin{cases}
	(2\pi)^{3/2}\delta(\vec{x}), & \alpha = 0,\\
\frac{2^{3/2+\alpha}\Gamma(\frac{3+\alpha}{2})}{\Gamma(-\frac{\alpha}{2})}
	\frac{1}{x^{3+\alpha}}, & \alpha \neq 0,
    \end{cases}
\end{equation*}
    so it makes sense to include into consideration their shifts
    in the coordinate space
\begin{equation*}
    \hat{v}_{\alpha, \vec{x}}(\vec{p})
	= e^{i\vec{p}\cdot\vec{x}} v_{\alpha}(\vec{p}).
\end{equation*}

    The Birman-Krein-Vishik theory
\cite{BKV}
    describes domain
    of extensions of the quadratic form
$ \lL $
    as the direct sum of
$ \DD(\lL) $
    and the linear span of the vector
$ R_{\mu} v $
\begin{equation}
\label{Dlk}
    \DD(\lL_{\kappa}) = \DD(\lL) \dotplus
    \{\xi R_{\mu} v\}, \quad \mu\in\CC\setminus\RR^{+}\cup 0 , \quad \xi \in
\CC.
\end{equation}
    Note that particular choice of
$ \mu $
    is irrelevant here since
\begin{equation*}
    R_{\mu_{1}}v - R_{\mu_{2}}v = (\mu_{1}-\mu_{2})
	R_{\mu_{1}} R_{\mu_{2}} v \in \DD(L),
    \quad \DD(L) \subset \DD(\lL).
\end{equation*}
    As shown in the work
\cite{KS}
    if the integral of
$ \hat{v}(\vec{p}) $
    in
(\ref{vcond2})
    is finite
    then the vector
$ R_{\mu}v $
    belongs to
$ \DD(\lL) $
    and we can define the action of
$ \lL_{\kappa} $
    as
\begin{equation}
\label{lk}
    \lL_{\kappa}[A] = \lL[A] + \kappa |(v,A)|^{2}
\end{equation}
    on the whole domain
$ \DD(\lL) $.
    That is 
$ \lL_{\kappa} $
    changes the action of
$ \lL $
    and so changes the action of
$ \HH $
    which is not our goal.
    Condition
(\ref{vcond2})
    breaks this picture in such a way that definition of
$ \lL_{\kappa} $
    now requires regularization of both terms
$ l[A] $
    and
$ |(v,A)|^{2} $
    and
    renormalization to zero of the running coefficient at the latter
\begin{equation}
\label{lkr}
    \lL_{\kappa}[A] = \lim_{\Lambda\to\infty}(\lL[P_{\Lambda}A]
	- \kappa_{\mathrm{r}}(\Lambda, \kappa) |(v,P_{\Lambda} A)|^{2}) ,
    \quad P_{\Lambda}\to \II, \quad \kappa_{\mathrm{r}} \to 0.
\end{equation}
    This way the second term tends to zero for 
$ A $ from
$ \DD(\lL) $
    and equation
(\ref{lext1})
    is restored.
    Meanwhile when acting on
$ R_{\mu} v $
    divergencies of
$ \lL[P_{\Lambda}R_{\mu}v] $
    and
$ \kappa_{\mathrm{r}}(\Lambda, \kappa) |(v,P_{\Lambda} R_{\mu}v)|^{2} $
    cancel each other and leave nontrivial difference in the limit
$ \Lambda\to\infty $.

    In the terms of potential
$ \hat{v}_{\alpha}(\vec{p}) $
    the divergence of the integral in
(\ref{vcond2})
    is expressed in the inequality
\begin{equation*}
    -\frac{1}{2} \leq \alpha .
\end{equation*}
    It is not hard to see that condition
(\ref{vcond1})
    imposes restriction on
$ \alpha $
    from the other side, namely the strict inequality
\begin{equation*}
    \alpha < \frac{1}{2} .
\end{equation*}
    While
$ \lL_{\kappa} $
    is an extension of the quadratic form
$ \lL $,
    the self-adjoint operator
$ L_{\kappa} $
    corresponding to 
$ \lL_{\kappa} $
    together with
$ L $
    are extensions of some other (symmetric) operator.
    This suggests to call
$ L_{\kappa} $
    not an extenstion but
    a singular perturbation of the operator
$ L $
    with 
$ v $
    playing the role of perturbation potential undergoing renormalization
\cite{AK},
\cite{Koshmanenko1}.

\subsection{Resolvent for interaction with single source}
    Domain of the self-adjoint operator
$ L_{\kappa} $
    corresponding to the quadratic form
$ \lL_{\kappa} $
    includes linear combinations of vectors
$ R_{\rho}v $,
$ R_{\rho}^{2} v $
    of the following type
\begin{equation*}
    R_{\rho} v + \xi(\kappa,\rho) R_{\rho}^{2} v, \quad \rho < 0 ,
	\quad \xi \in \CC,
\end{equation*}
    with some
$ \xi $
    defined by the boundary conditions of
$ L_{\kappa} $.
    However we will not stop on this domain,
    neither we will not study the renormalization in
(\ref{lkr}),
    since the most convenient object for construction of the square root of
$ L_{\kappa} $
    is its resolvent given by the Krein formula
\cite{Krein}
\begin{equation}
\label{resK}
    R_{\mu}^{\kappa} = (L_{\kappa}-\mu \II)^{-1}
    = R_{\mu} +
	\frac{R_{\mu}v (R_{\bar{\mu}}v, \,\cdot\,)}{\kappa - \sigmab(\mu)}.
\end{equation}
    Here
$ \sigmab(\mu) $
    is the Nevanlinna (Herglotz) function
\cite{GT},
    defined by the equation
\begin{equation}
\label{sigmaDef}
    \sigmab(\mu) - \sigmab(\nu) = (\mu-\nu) (R_{\bar{\mu}}v, R_{\nu}v)
	= (v, (R_{\mu}-R_{\nu})v)
\end{equation}
    and
$ \kappa $
    is the real separation constant of the particular solution
    which has the meaning of
    extension parameter.
    Note that in the case
$ \kappa = \infty $
    the second term in the RHS of
(\ref{resK})
    vanishes and we arrive at the resolvent
$ R_{\mu} $
    of
$ L $
\begin{equation*}
    R_{\mu}^{\infty} = R_{\mu} .
\end{equation*}
    In what follows we will restrict our consideration to the potential
\begin{equation*}
\hat{v}(\vec{p}) = \hat{v}_{-1/2}(\vec{p}) = \frac{1}{p^{1/2}}
\end{equation*}
    and its coordinate shifts,
    such that vectors
$ R_{\mu}v $
    reside in the region described by
(\ref{ABound}).
    Indeed,
$ R_{\mu}v $
    in the coordinate representation is given by the integral
\begin{equation*}
    (R_{\mu}v)(\vec{x}) = \frac{1}{(2\pi)^{3/2}} \int_{\RR^{3}}
	\frac{e^{i\vec{p}\cdot\vec{x}}\, d^{3}p}{(p^{2}-\mu) p^{1/2}}
    = \frac{4\pi}{(2\pi)^{3/2}x} \int_{0}^{\infty} \frac{\sin px}{p^{2}-\mu}
    p^{1/2}\, dp,
\end{equation*}
    which after the change of variable
$ s=px $ 
    turns into
\begin{equation}
\label{Rvx}
    (R_{\mu}v)(\vec{x})
    = \frac{4\pi}{(2\pi)^{3/2}x^{1/2}}
	\int_{0}^{\infty} \frac{\sin s}{s^{2}-\mu x^{2}} s^{1/2}\, ds .
\end{equation}
    The integral by
$ s $
    converges absolutely, it is bounded for
$ x \to 0 $
    and it is not equal to zero in this limit. Then one may conclude that
    singularity of
$ (R_{\mu}v)(\vec{x}) $
    is defined by the coefficient
$ x^{-1/2} $
    and so is described by
(\ref{ABound}).

    The Nevanlinna function
    for
$ \hat{v}(\vec{p})=p^{-1/2} $
    can be derived from the following equation
\begin{equation}
\label{Nln}
    \sigmab(\mu) - \sigmab(\lambda) = \int_{\RR^{3}}\bigl(
	\frac{1}{p^{2}-\mu}-\frac{1}{p^{2}-\lambda}\bigr) \frac{d^{3}p}{p}
    = 2\pi (\ln(-\lambda) - \ln(-\mu)).
\end{equation}
    Here we choose the main branch of the logarithm with the cut along
    the negative semi-axis in such a way that RHS 
    is defined for all complex
$ \mu $ and $ \lambda $
    outside of the spectrum of
$ L $.
    This way the denominator in 
(\ref{resK})
    can be chosen as
\begin{equation}
\label{gln}
    \kappa - \sigmab(\mu) = 2\pi \ln(-\frac{\mu}{\tilde{\kappa}}),
    \quad \tilde{\kappa} > 0,
\end{equation}
    where
$ \tilde{\kappa} $
    is a positive value combining dimensional constant to correctly
    define the logarithm and the extension parameter
$ \kappa $.

    As long as the logarithm is an injective function 
$ \CC\setminus\{\RR^{-}\cup 0\} \to \CC $,
    the resolvent
(\ref{resK})
    for any real
$ \kappa $
    (or positive
$ \tilde{\kappa} $)
    has exactly one pole located in
$ \mu=-\tilde{\kappa} $.
    And the residue of this pole is the Riesz projection
    on the negative eigenspace of the operator
$ L_{\kappa} $.
    This contradicts the requirement of positivity of the extension
$ \lL_{\kappa} $
    and shows that nontrivial solution to the eigenstate equation of 
$ \HH $
    produced by a single localized external source does not exist
    in our setup.

    It is worth to note that logarithmic Nevanlinna function
(\ref{Nln})
    and analysis of extensions of the quadratic form coincide
    with the corresponding objects of the theory of singular perturbations
    of 2-dimensional Laplace operator by the
$ \delta $-potential
    described in
\cite{LFres}.

\subsection{Interaction with two sources}
    In order to fix obstacles encountered in the previous section
    let us turn to a more complex singular potential.
    In the original work
\cite{Krein}
    Krein equation
    is derived for the interaction with multiple sources
$ v_{m} $
\begin{equation*}
    R^{\kappa}_{\mu} = R_{\mu} + R_{\mu}v_{m} (\kappa-\sigmab(\mu))_{ml}^{-1}
	(R_{\bar{\mu}}v_{l}, \,\cdot\,),
\end{equation*}
    where
$ \kappa $ 
    is now a real symmetric matrix and the matrix
$ \sigmab $
    is defined by the equation
\begin{equation}
\label{NF}
    \sigmab_{jk}(\mu) - \sigmab_{jk}(\nu) = (v_{j}, (R_{\mu}-R_{\nu})v_{k}).
\end{equation}
    In our case
    it is natural to add to the resolvent the potential
\begin{equation*}
    \hat{v}_{\vec{x}}(\vec{p}) = e^{i\vec{p}\cdot\vec{x}} p^{-1/2}
\end{equation*}
    shifted by the vector
$ \vec{x} $
    in the coordinate space.
    In order to avoid the treatment of indices of
    the matrix
$ \kappa_{jk} - \sigmab_{jk}(\lambda) $
    one may turn to the basis
\begin{align*}
    \hat{v}_{-}(\vec{p}) &=
    \frac{1}{\sqrt{2}}(\hat{v}(\vec{p}) - \hat{v}_{\vec{x}}(\vec{p}))
	= \frac{1}{\sqrt{2}} p^{-1/2}(1-e^{i\vec{p}\cdot\vec{x}}) ,\\
    \hat{v}_{+}(\vec{p}) &=
    \frac{1}{\sqrt{2}}(\hat{v}(\vec{p}) + \hat{v}_{\vec{x}}(\vec{p}))
	= \frac{1}{\sqrt{2}} p^{-1/2}(1+e^{i\vec{p}\cdot\vec{x}})
\end{align*}
    and restrict consideration to Krein resolvents with nontrivial second term
    generated by the antisymmetric potential
$ v_{-}(\vec{p}) $
\begin{equation}
\label{resKminus}
    R_{\mu}^{\kappa} = R_{\mu} +
	\frac{R_{\mu}v_{-} (R_{\bar{\mu}}v_{-}, \,\cdot\,)}{\kappa
	    - \sigmab(\mu)}.
\end{equation}
    In this case
    we have the following equation for
$ \sigmab(\mu) $
\begin{multline}
    \sigmab(\mu) - \sigmab(\lambda) = (v_{-}, (R_{\mu}- R_{\lambda})v_{-}) =\\
    = \int_{\RR^{3}}
	\frac{1-\cos(\vec{p}\cdot\vec{x})}{p}
	\bigl(\frac{1}{p^{2}-\mu}
	-\frac{1}{p^{2}-\lambda}\bigr) \,
	d^{3}p =\\
\label{gmuminus}
    = 4\pi \int_{0}^{\infty}\bigl(p - \frac{\sin px}{x}\bigr)\bigl(
	\frac{1}{p^{2}-\mu} - \frac{1}{p^{2}-\lambda}
    \bigr) dp .
\end{multline}
    Similarly to
(\ref{gln})
    let us introduce the dimensional parameter
$ \tilde{\kappa} $
    which includes
$ \kappa $
    and define denominator in
(\ref{resKminus})
    by means of the integral
\begin{multline}
    \sigmab(\mu)-\kappa = 4\pi\int_{0}^{\infty} \bigl(
	\frac{1}{p^{2}-\mu} \bigl(p - \frac{\sin px}{x} \bigr)
	- \frac{p}{p^{2}+\tilde{\kappa}}
    \bigr) dp =\\
\label{gmk}
    = 2\pi\ln\frac{\tilde{\kappa}}{-\mu}
	-4\pi \int_{0}^{\infty} \frac{\sin px}{x} \frac{dp}{p^{2}-\mu} .
\end{multline}
    The integral in the first line, being a sum of elementary Nevanlinna
    functions
\begin{equation*}
	\frac{1}{p^{2}-\mu} \bigl(p - \frac{\sin px}{x} \bigr)
	- \frac{p}{p^{2}+\tilde{\kappa}}, \quad p - \frac{\sin px}{x} \geq 0,
\end{equation*}
    is Nevanlinna function itself,
    that is a holomorphic map of the upper half-plane into
    (a part of) itslef. From this it immediately follows that preimage of zero
    for
$ \sigmab(\mu) $
    is either undefined or located on the boundary of the upper half-plane.
    The preimage (or preimages) of zero also cannot be
    positive as long as in this case the first integral in
(\ref{gmk})
    has non-trivial imaginary part generated by the vicinity of
    the point
$ p = \sqrt{\mu} $,
    while the rest of the integral is real.
    This way all zeros of
(\ref{gmk})
    are located on the negative semi-axis and our goal is to study the
    behaviour of this expression thereon.
    Its second term is a function of the product
$ \mu x^{2} $
    and vanishes for
$ \mu\to -\infty $
\begin{equation}
\label{oscexpinfty}
    \JJ(\mu x^{2})
    = 4\pi \int_{0}^{\infty} \frac{\sin px}{x} \frac{dp}{p^{2}-\mu}
	\stackrel{\mu x^{2}\to -\infty}{=}
	- \frac{4\pi}{\mu x^{2}} + \frac{8\pi}{\mu^{2}x^{4}}
	+ \OO(\frac{1}{\mu^{3} x^{6}}) .
\end{equation}
    Thus the whole sum
(\ref{gmk})
    goes logarithmically to
$ -\infty $
\begin{equation*}
    \sigmab(\mu) - \kappa = 2\pi\ln\frac{\tilde{\kappa}}{-\mu}
	+ \frac{4\pi}{\mu x^{2}} + \OO(\mu^{-2}) 
	\to -\infty, \quad \mu\to -\infty.
\end{equation*}
    Derivative of the function
$ \sigmab(\mu) $
    for
$ \mu < 0 $
    is strictly positive
\begin{equation}
\label{sigmader}
    \sigmab'(\mu) = 4\pi \int_{0}^{\infty} \bigl(p- \frac{\sin px}{x}\bigr)
	\frac{dp}{(p^{2}-\mu)^{2}} > 0,
\end{equation}
    which means that
$ \sigmab(\mu) $
    monotonously increases with the increase of
$ \mu $.
    In order to estimate the value 
$ \sigmab(0) $
    let us write the integral
$ \JJ(\mu x^{2}) $
    for finite
$ \mu $ and
$ x $
    in the terms of special functions
\begin{align}
\label{intSC1}
    4\pi \int_{0}^{\infty} & \frac{\sin px}{x} \frac{dp}{p^{2}-\mu}
	= \frac{4\pi}{\sqrt{\mu}x} \bigl(
	\cos\sqrt{\mu}x \, \Si\sqrt{\mu}x -\sin\sqrt{\mu}x \,\Ci\sqrt{\mu} x
    \bigr) = \\
\label{intSC2}
    & \quad \stackrel{\mu x^{2}\to 0}{=}
  4\pi (1- \boldsymbol{\gamma} - \frac{1}{2}\ln(-\mu x^{2}) 
    + \frac{1}{12}\mu x^{2} \ln(-\mu x^{2})
	+ \OO(\mu x^{2})).
\end{align}
    Here integral sine
$ \Si(\sqrt{\mu}x) $
    and cosine
$ \Ci(\sqrt{\mu}x) $
    with ``symmetric'' imaginary part
    are defined as
\begin{gather}
\label{Si}
    \Si(\sqrt{\mu}x) = \int_{0}^{\sqrt{\mu}x} \frac{\sin t}{t} dt, \\
\label{Ci}
    \Ci(\sqrt{\mu}x) = \boldsymbol{\gamma} + \frac{1}{2}\ln (-\mu x^{2})
	- \int_{0}^{\sqrt{\mu}x} \frac{1-\cos t}{t} dt,
\end{gather}
    and
$ \boldsymbol{\gamma} $
    is the Euler constant.
    One may note that RHS in
(\ref{intSC1})
    includes only Taylor-expandable even functions and the logarithm.
    That is branching by
$ \mu $
    in
(\ref{intSC1})
    is defined solely by the logarithm from the integral cosine
(\ref{Ci}).
    Substituting
(\ref{intSC2})
    into
(\ref{gmk})
    one derives the following expansion for
$ \sigmab(\mu) - \kappa $
\begin{equation}
\label{gmuzero}
    \sigmab(\mu) - \kappa = 2\pi\ln\tilde{\kappa} x^{2}
	+ 4\pi(\boldsymbol{\gamma} - 1) 
    -\frac{\pi}{3}\mu x^{2} \ln (-\mu x^{2}) + \OO(\mu x^{2}) ,
    \quad \mu \to 0.
\end{equation}
    This equation shows that we can always choose
$ \tilde{\kappa} $
    in such a way that RHS of
(\ref{gmuzero})
    is negative when
$ \mu $
    tends to zero.
    The latter means, due to the positivity of derivative
(\ref{sigmader}),
    that
$ \sigmab(\mu)-\kappa $
    does not turn to zero on the negative semi-axis and that for such
$ \tilde{\kappa} $
    resolvent
(\ref{resKminus})
    defines positive operator
$ L_{\kappa} $
    and thus real-valued quadratic form
$ \omega_{\kappa}[A] $.

\begin{figure}
  \includegraphics[width=\linewidth]{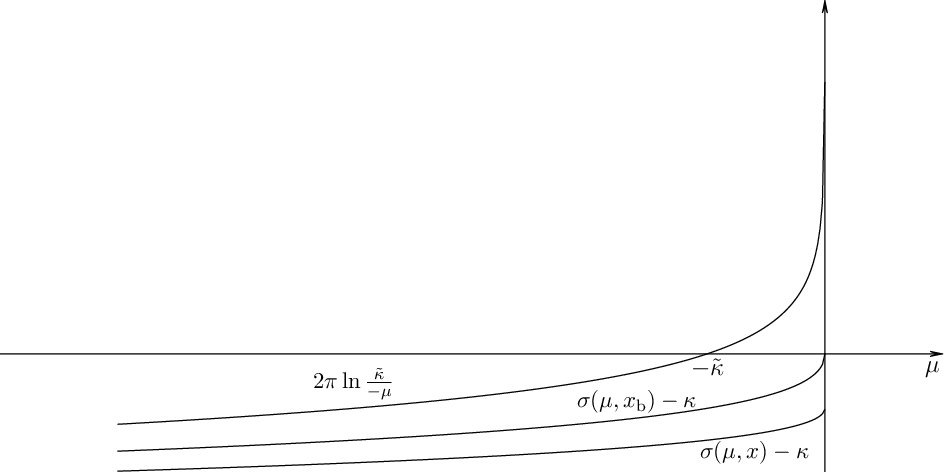}
  \caption{Functions $\sigmab(\mu, x)$ (bottom),
    $\sigmab(\mu, x_{\mathrm{b}})$ (middle) compared
    to $2\pi \ln \frac{\tilde{\kappa}}{-\mu}$}
  \label{fig:sigma}
\end{figure}

    Let us look at
(\ref{gmuzero})
    from the other side. Suppose that for chosen values of
$ \tilde{\kappa} $ and
$ x $
    the limit of RHS in
(\ref{gmuzero})
    is negative and we can correctly define the functional
$ \Omega_{\kappa}(A) $.
    When increasing the distance
$ x $
    the limit of RHS in
(\ref{gmuzero})
    is increasing as well and for some boundary
$ x=x_{\mathrm{b}} $
    it turns to zero (see Figure~\ref{fig:sigma})
\begin{equation}
\label{sigmazero}
    \sigmab(\mu, x_{\mathrm{b}})|_{\mu = 0} - \kappa = 0.
\end{equation}
    Now
$ \mu = 0 $
    is not a pole, but rather a branch point, as can be seen from
(\ref{gmuzero})
    -- the first two terms cancel each other while the product
$ \mu x^{2} \ln(-\mu x^{2}) $
    from the third one tends to zero slower than
$ \mu $.

    For
$ x > x_{\mathrm{b}} $
    the function
$ \sigmab(\mu, x)-\kappa $
    changes its sign on the interval
$ -\infty < \mu < 0 $
    and the corresponding self-adjoint operator
$ L_{\kappa} $
    possesses negative eigenvalue.
    That is for any chosen extension parameter
$ \tilde{\kappa} $
    admissible relative positions of sources
$ \vec{x}_{1} $
    and
$ \vec{x}_{2} $
    always exist
    and are always limited by the sphere with radius proportional to the
    inverse square root of
$ \tilde{\kappa} $
\begin{equation*}
    |\vec{x}_{2} - \vec{x}_{1}| < x_{\mathrm{b}}
	= \frac{e^{1-\boldsymbol{\gamma}}}{\sqrt{\tilde{\kappa}}} .
\end{equation*}
    Evidently extensions of the quadratic form of 
    Laplacian by potentials
    with singularities in three or more points are also possible, but
    in this case the
    analysis of admissible positions of singularities
    is more complex.

\subsection{Electromagnetic field in the Coulomb gauge}
    Quantum Hamiltonian of the free electromagnetic field in the Coulomb gauge
    acts as the following operator
\cite{Hatfield}
\begin{equation}
\label{HEM}
    \HH  = \int_{\RR^{3}}d^{3}p \, \bigl(
    -\frac{\delta}{\delta \hat{A}_{l}(-\vec{p})} \hat{P}_{lj}^{\bot}
	\frac{\delta}{\delta \hat{A}_{j}(\vec{p})}
+ p^{2} |\hat{A}_{l}(\vec{p})|^{2}
    \bigr) ,
\end{equation}
    where
$ \hat{P}_{lj}^{\bot} $
    is the projector to the transverse (solenoidal) component
\begin{equation*}
    \hat{P}_{lj}^{\bot} = \delta_{lj} - \frac{p_{l}p_{j}}{p^{2}} .
\end{equation*}
    Corresponding functional of the free ground state 
    differs from its scalar counterpart
    only by summation of indices in the quadratic form
$ \omega[A] $
\begin{equation*}
    \Omega_{0}(A) = \exp\{-\frac{1}{2}\omega[A]\}, \quad
    \omega[A] = \int_{\RR^{3}} p |\hat{A}_{l}(\vec{p})|^{2} \, d^{3}p ,
\end{equation*}
    provided that
$ \hat{A}_{l}(\vec{p}) $
    is transverse
\begin{equation*}
    p_{l} \hat{A}_{l}(\vec{p}) = 0.
\end{equation*}
    The latter condition ensures that linear functional (1-form)
\begin{equation*}
    \frac{\delta \omega[A]}{\delta \hat{A}_{l}(\vec{p})}
\end{equation*}
    is transverse \textit{w.\,r.}~to variables
$ (\vec{p}, l) $
    as well,
    and that the projector
$ \hat{P}_{lj}^{\bot} $
    in the middle of the first term in
(\ref{HEM})
    acts as unity when
$ \HH $
    is applied to
$ \Omega_{0}(A) $.

    In order to construct closed positive extensions of the quadratic form
\begin{equation*}
    \lL[A] = \int_{\RR^{3}} p^{2} |\hat{A}_{l}(\vec{p})|^{2} \, d^{3}p
\end{equation*}
    from the second term of
$ \HH $
    let us consider the following vector singular potentials
\begin{equation*}
    \hat{\vec{v}}^{n} = \frac{\vec{p}\times(\vec{x}_{2}-\vec{x}_{1})}{
	p^{3/2}|\vec{x}_{2}-\vec{x}_{1}|}e^{i\vec{p}\cdot\vec{x}_{n}},\quad
    \hat{\vec{u}}^{n}
	= \frac{\vec{p}\times(\vec{p}\times(\vec{x}_{2}-\vec{x}_{1}))}{
	p^{5/2}|\vec{x}_{2}-\vec{x}_{1}|}e^{i\vec{p}\cdot\vec{x}_{n}},\quad
    n=1,2,
\end{equation*}
    attached to the points
$ \vec{x}_{1} $
    and
$ \vec{x}_{2} $.
    These potentials are transverse and
$ \vec{v}^{1,2} $
    are orthogonal to
$ \vec{u}^{1,2} $
    as 3-dimensional vectors,
    which means that
$ 4\times 4 $
    matrix in the RHS of
(\ref{NF})
    separates into two
$ 2\times 2 $
    matrices
\begin{align}
\label{smu}
    \sigmab_{mn}^{v}(\mu) - \sigmab_{mn}^{v}(\nu)
	& = (v_{l}^{m}, (R_{\mu}-R_{\nu})v_{l}^{n}), \\
\label{snu}
    \sigmab_{mn}^{u}(\mu) - \sigmab_{mn}^{u}(\nu)
	& = (u_{l}^{m}, (R_{\mu}-R_{\nu})u_{l}^{n}), \\
\nonumber
    (u_{l}^{m}, (R_{\mu}-R_{\nu})v_{l}^{n})
	&= (v_{l}^{m}, (R_{\mu}-R_{\nu})u_{l}^{n}) = 0. 
\end{align}
    Straightforward calculation similar to
(\ref{gmuminus}),
(\ref{gmk})
    shows that matrices
$ \sigma_{mn}^{v,u} $
    with off-diagonal separation constants set to zero are equal
    to each other and have the following form
\begin{equation*}
    \sigmab_{mn}^{v,u}(\mu) =
\begin{pmatrix}
    \frac{4\pi}{3} \ln \frac{\tilde{\kappa}}{-\mu} &
    -\frac{8\pi}{\mu x^{2}} - 4\JJ'(\mu x^{2}) \\
    -\frac{8\pi}{\mu x^{2}} - 4\JJ'(\mu x^{2}) &
    \frac{4\pi}{3} \ln \frac{\tilde{\kappa}}{-\mu} .
\end{pmatrix},
    \quad x = |\vec{x}_{2}-\vec{x}_{1}| ,
\end{equation*}
    This equation
    reproduces the result in
\cite{TBZ}
    however for slightely different transverse vector potentials.
    Matrices
$ \sigmab_{mn}^{v,u} $
    can be trivially diagonalized by turning to symmetric and antisymmetric
    combinations of potentials for the points
$ \vec{x}_{1} $ and
$ \vec{x}_{2} $,
    the result is represented by the following Nevanlinna functions
\begin{align}
\label{sigmavup}
    \sigmab^{v,u}_{+}(\mu) &= \frac{4\pi}{3}\ln \frac{\tilde{\kappa}}{-\mu}
	-\frac{8\pi}{\mu x^{2}} - 4\JJ'(\mu x^{2}), \\
\label{sigmavum}
    \sigmab^{v,u}_{-}(\mu) &= \frac{4\pi}{3}\ln \frac{\tilde{\kappa}}{-\mu}
	+\frac{8\pi}{\mu x^{2}} + 4\JJ'(\mu x^{2}) .
\end{align}
    Calculation of higher terms in 
(\ref{intSC2})
    yields the following expansion of
$ \JJ'(\mu x^{2}) $
    for the small argument
\begin{equation*}
    \JJ'(\mu x^{2}) \stackrel{\mu x^{2}\to 0}{=}
    2\pi\bigl(-\frac{1}{\mu x^{2}}
	+\frac{1}{6}\ln(-\mu x^{2}) + \frac{\boldsymbol{\gamma}}{3}
	- \frac{4}{9} + \ln(-\mu x^{2})\OO(\mu x^{2})\bigr).
\end{equation*}
    This allows to estimate behaviour of 
$ \sigmab^{v,u}_{+}(\mu) $,
$ \sigmab^{v,u}_{-}(\mu) $
    on the interval 
$ -\infty < \mu < 0 $
    and find out that the RHS of
(\ref{sigmavup})
    changes from
$ -\infty $
    to
$ +\infty $
    thereon, while the RHS of
(\ref{sigmavum})
    is upper-bounded. 
    Thus only 
$ \sigmab^{v,u}_{-}(\mu) $
    are zeroless for corresponding values of
$ \tilde{\kappa} $
    and can be used to construct positive extensions of the quadratic form in
(\ref{HEM}).

\section{Square root of the operator $ L_{\kappa} $}
\subsection{Dunford integral for the square root}
    The square root of the positive self-adjoined operator
$ L_{\kappa} $
    can be written by means of the Riesz spectral decomposition
\cite{Riesz}
    which coincide with the Dunford integral
\cite{DS}
    of the resolvent
$ R_{\lambda}^{\kappa} $
    around the positive semi-axis in the clockwise direction
\begin{equation*}
    L_{\kappa}^{1/2} = \lim_{\epsilon\to 0} \frac{1}{2\pi i}
	\int_{0}^{\infty}(
    R_{\lambda+i\epsilon}^{\kappa} - R_{\lambda-i\epsilon}^{\kappa})
	\lambda^{1/2} \, d\lambda .
\end{equation*}
    Substituting the expression for
$ R_{\lambda}^{\kappa} $
    from
(\ref{resK})
    and using the notation
$ \pm i0 $
    for the limit in
$ \epsilon $
    one can write the formal representation
\begin{multline}
    L_{\kappa}^{1/2} = \frac{1}{2\pi i}
	\int_{0}^{\infty}\bigl(
    R_{\lambda+i0}- R_{\lambda-i0} + \\
    + \frac{R_{\lambda+i0}v (R_{\lambda-i0}v, \, \cdot \,)}{
	\kappa - \sigmab(\lambda +i0)}
    - \frac{R_{\lambda-i0}v (R_{\lambda+i0}v, \, \cdot \,)}{
	\kappa - \sigmab(\lambda -i0)}
    \bigr) \lambda^{1/2} \, d\lambda = \\
\label{Lkappa12}
    = p + \frac{1}{2\pi i}
	\int_{0}^{\infty}\bigl(
    \frac{R_{\lambda+i0}v (R_{\lambda-i0}v, \, \cdot \,)}{
	\kappa - \sigmab(\lambda +i0)}
    - \frac{R_{\lambda-i0}v (R_{\lambda+i0}v, \, \cdot \,)}{
	\kappa - \sigmab(\lambda -i0)}
    \bigr) \lambda^{1/2} \, d\lambda ,
\end{multline}
    where
$ p $
    is the operation of multiplication by the modulus of momentum
    variable 
$ \vec{p} $.
    In particluar
$ p $
    is the action of the square root of Laplacian
\begin{equation}
\label{L12}
    L^{1/2} :\ \hat{A}(\vec{p}) \to p \hat{A}(\vec{p}),
\end{equation}
    but we cannot substitute this in
(\ref{Lkappa12})
    as long as
$ L_{\kappa}^{1/2} $ and
$ L^{1/2} $
    are possibly defined on different domains. Nevertheless the expression
(\ref{Lkappa12}),
    in general case with some regularization,
    can be used as the kernel of the quadratic form of
$ L_{\kappa}^{1/2} $
    and for calculation of some traces.
    For this purpose we will also need the gap of
$ \sigmab(\lambda) $
    across the positive semi-axis, which in the scalar case is equal to
\begin{multline}
    \sigmab(\lambda+i0) -\sigmab(\lambda-i0) =
	2\pi\ln\frac{\tilde{\kappa}}{-\lambda-i0}
	-2\pi\ln\frac{\tilde{\kappa}}{-\lambda+i0} -\\
\label{gap1}
    -4\pi\int_{0}^{\infty}\frac{\sin px}{x}(\frac{1}{p^{2}-\lambda -i0}
	- \frac{1}{p^{2}-\lambda +i0}) \,dp
    = 4\pi^{2} i - 4\pi^{2}i \frac{\sin \sqrt{\lambda} x}{\sqrt{\lambda}x} ,
\end{multline}
    and for extensions of the quadratic form of Coulomb-gauged
    vector Laplacian it is
\begin{equation}
\label{gap2}
    \sigmab^{v,u}_{-}(\lambda+i0) -\sigmab^{v,u}_{-}(\lambda-i0) =
    \frac{8\pi^{2}i}{3} +
	8\pi^{2}i\bigl(
	    \frac{\cos\sqrt{\lambda}x}{\lambda x^{2}}
	    - \frac{\sin\sqrt{\lambda}x}{(\sqrt{\lambda}x)^{3}}
	      \bigr) .
\end{equation}
    One can notice that real coefficients at
$ \mu $
    in functions
$ R_{\mu} $
    and
$ \sigmab(\mu) $
    yield relations
\begin{equation*}
    \ol{R_{\mu}} = R_{\bar{\mu}}, \quad
    \ol{\sigmab(\mu)} = \sigmab(\bar{\mu}),
\end{equation*}
    and ensure that the kernel of
$ L_{\kappa}^{1/2} $ from
(\ref{Lkappa12})
    in the coordinate representation
    is real and symmetric,
    as is required by the equation
(\ref{lqeq}).

    The square root
(\ref{Lkappa12})
    can also be used to establish asymptotic freedom of functional
(\ref{OmegaK}).
    By this notion we mean that density of expectation given
    by the eigenstate
$ \Omega_{\kappa}(A) $
    tends to that of the eigenstate
$ \Omega_{0}(A) $
    when we either cut off frequencies lower than
$ \Lambda $ 
    with the Heaviside step
\begin{equation*}
    \frac{\Omega_{\kappa}(\chi_{\Lambda}A)}{\Omega_{0}(\chi_{\Lambda}A)}
	\stackrel{\Lambda \to \infty}{\to} 1, 
    \quad \chi_{\Lambda}: \ \hat{A}(\vec{p}) \to
	\begin{cases} 0, & |\vec{p}| < \Lambda \\
	    \hat{A}(\vec{p}), & |\vec{p}| > \Lambda
	\end{cases}, \quad A\in\DD(\lL)
\end{equation*}
    or scale
$ \hat{A}(\vec{p}) $
    off the center to the UV-region
\begin{equation*}
    \frac{\Omega_{\kappa}(S_{\rho} A)}{
	\Omega_{0}(S_{\rho} A)} \stackrel{\rho\to 0}{\to} 1,
	\quad S_{\rho} : \ \hat{A}(\vec{p}) \to \rho \hat{A}(\rho\vec{p}),
    \quad A\in\DD(\lL).
\end{equation*}
    We will not stop on these statements as they are beyond the scope
    of the present work.

\subsection{Reconstruction of the quadratic form of
$ L_{\kappa}^{1/2} $}
    As was stated in the Introduction, in order to verify that
$ \Omega_{\kappa} $
    is an eigenfunctional of the operation
$ \HH $,
    one needs to prove that the positive quadratic form of
$ L_{\kappa}^{1/2} $
    can be reconstructed by its action on the set
$ \DD(\lL) $
    everywhere dense in the space
$ L_{2}(\RR^{3}) $.
    Let
$ A \in \DD(\omega_{\kappa}) $,
    then there exists a sequence
$ A_{n} \in \DD(L_{\kappa}^{1/2}) $
    converging to
$ A $
    in the norm induced by the form
$ \omega_{\kappa} $
\begin{equation*}
    || A - A_{n} ||_{L_{2}(\RR^{3})} \to 0 ,\quad
    \omega_{\kappa}[A-A_{n}] \to 0, \quad n \to \infty.
\end{equation*}
    Thus it is sufficient to prove that any function
$ A_{n} \in \DD(L_{\kappa}^{1/2}) $
    can be approximated in the norm of
$ \omega_{\kappa} $
    by functions from
$ \DD(\lL) $.
    If
$ A_{n} $
    belongs to
$ \DD(L_{\kappa}^{1/2}) $
    then it belongs to the domain of quadratic form
$ \lL_{\kappa} $
    of the operator
$ L_{\kappa} $
\begin{equation*}
    \lL_{\kappa}[A_{n}] =
    (L_{\kappa}^{1/2} A_{n}, L_{\kappa}^{1/2} A_{n}) < \infty .
\end{equation*}
    As long as
$ \DD(\lL_{\kappa}) $
    is the direct sum of
$ \DD(\lL) $
    and the linear span of some vector
$ R_{\rho}v $,
$ \rho < 0 $,
    then this
$ A_{n} $
    can be uniquely decomposed into the sum
\begin{equation*}
    A_{n} = a_{n} + \xi_{n} R_{\rho} v, \quad a_{n}\in\DD(\lL),
	\quad \xi_{n} \in \CC, \quad \rho < 0.
\end{equation*}
    By this reason it is sufficient to show that just the function
$ R_{\rho} v $
    can be approximated by functions from
$ \DD(\lL) $
    in the norm of the quadratic form
$ \omega_{\kappa} $.

    In order to verify the above proposition let us use expression
(\ref{Lkappa12})
    for the operator
$ L_{\kappa}^{1/2} $
    and write down its quadratic form
\begin{multline}
    \omega_{\kappa}[R_{\rho}v] =
    \omega[R_{\rho}v] + \frac{1}{2\pi i}\int_{0}^{\infty} \bigl(
\frac{(R_{\rho}v,R_{\lambda+i0}v)(R_{\lambda-i0}v,R_{\rho}v)}{
    \kappa - \sigmab(\lambda+i0)} - \\
\label{omegaRv}
-
\frac{(R_{\rho}v,R_{\lambda-i0}v)(R_{\lambda+i0}v,R_{\rho}v)}{
    \kappa - \sigmab(\lambda-i0)}
    \bigr)\lambda^{1/2} \, d\lambda.
\end{multline}
    Now using
(\ref{sigmaDef})
    each term in the integrand can be transformed into 
\begin{multline*}
\frac{(R_{\rho}v,R_{\lambda\pm i0}v)
    (R_{\lambda\mp i0}v,R_{\rho}v)}{\kappa - \sigmab(\lambda\pm i0)}	
    = \frac{(\sigmab(\lambda\pm i0) - \sigmab(\rho))^{2}}{
    (\lambda \pm i0 - \rho)^{2}(\kappa - \sigmab(\lambda\pm i0))} =\\
    = \frac{2\sigmab(\rho)-\kappa -\sigmab(\lambda\pm i0)}{
    (\lambda \pm i0 - \rho)^{2}}
    + \frac{(\kappa-\sigmab(\rho))^{2}}{
    (\lambda \pm i0 - \rho)^{2}(\kappa - \sigmab(\lambda\pm i0))} ,
\end{multline*}
    and finally, 
    up to some vanishing terms proportional to
$ i0 $,
    we have
\begin{multline}
    \omega_{\kappa}[R_{\rho}v] = 4\pi\int_{0}^{\infty}
	\frac{p^{2}\, dp}{(p^{2}-\rho)^{2}}
    - \frac{1}{2\pi i}
    \int_{0}^{\infty} 
    \frac{\sigmab(\lambda+ i0) - \sigmab(\lambda- i0)}{
    (\lambda - \rho)^{2}} \lambda^{1/2} d\lambda \\
    + \frac{1}{2\pi i} (\kappa-\sigmab(\rho))^{2}
    \int_{0}^{\infty} 
    \frac{\sigmab(\lambda+ i0) - \sigmab(\lambda- i0)}{
    (\lambda - \rho)^{2}
	(\kappa - \sigmab(\lambda+i0))(\kappa - \sigmab(\lambda-i0))
    } \lambda^{1/2} d\lambda .
\label{wkRv}
\end{multline}
    The modulus of the gap of
$ \sigmab(\lambda) $
    calculated in
(\ref{gap1})
    is limited, thus all three integrals in
(\ref{wkRv})
    converge absolutely.
    Then we may approximate
$ R_{\rho}v $
    by the function
$ (\Lambda+\rho)R_{-\Lambda}R_{\rho}v $
    from
$ \DD(\lL) $
    using the trivial identity
\begin{equation*}
    R_{\rho}v = (\Lambda +\rho) R_{-\Lambda} R_{\rho} v + R_{-\Lambda} v
\end{equation*}
    and the following limit in the norm
$ L_{2}(\RR^{3}) $
\begin{equation*}
    ||R_{\rho}v - (\Lambda +\rho) R_{-\Lambda} R_{\rho} v ||
	= || R_{-\Lambda} v || = ||\frac{1}{(p^{2}+\Lambda)p^{1/2}}||
    \stackrel{\Lambda\to\infty}{\to} 0 .
\end{equation*}
    The same limit is valid in the norm of the form
$ \omega $
\begin{equation*}
    \omega[R_{-\Lambda}v] = \int_{\RR_{3}}
	\frac{d^{3}p}{(p^{2}+\Lambda)^{2}} \to 0, \quad \Lambda \to \infty ,
\end{equation*}
    thus in order to prove that
\begin{equation*}
    \omega_{\kappa}[R_{\rho}v - (\Lambda +\rho) R_{-\Lambda} R_{\rho} v]
    = \omega_{\kappa}[R_{-\Lambda}v] \to 0, \quad \Lambda \to \infty ,
\end{equation*}
    one needs to show that the second and the third terms in
(\ref{wkRv}),
    after the substitution of
$ \rho $
    with
$ -\Lambda $,
    tend to zero for
$ \Lambda \to \infty $.
    By the change of variable
$ \tilde{\lambda} = \lambda/\Lambda $
    those terms can be rewritten as
\begin{gather}
\label{L1}
    - \frac{1}{2\pi i \Lambda^{1/2}}
    \int_{0}^{\infty} 
    \frac{\sigmab(\tilde{\lambda}\Lambda+ i0)
	- \sigmab(\tilde{\lambda}\Lambda- i0)}{
    (\tilde{\lambda} + 1)^{2}} \tilde{\lambda}^{1/2} d\tilde{\lambda} \\
\label{L2}
    \frac{(\kappa-\sigmab(-\Lambda))^{2}}{2\pi i \Lambda^{1/2}} 
    \int_{0}^{\infty} 
    \frac{\sigmab(\tilde{\lambda}\Lambda+ i0)
	- \sigmab(\tilde{\lambda}\Lambda- i0)}{
    (\tilde{\lambda} +1)^{2}
	(\kappa - \sigmab(\tilde{\lambda}\Lambda+i0))
	(\kappa - \sigmab(\tilde{\lambda}\Lambda-i0))
    } \tilde{\lambda}^{1/2} d\tilde{\lambda} ,
\end{gather}
    correspondingly.
    Here both integrals converge absolutely and due to the limited gap of
$ \sigmab(\lambda) $
    their values are bounded from above for any
$ \Lambda $.
    Taking into account coefficients
$ \Lambda^{-1/2} $
    one may conclude that
(\ref{L1}),
(\ref{L2})
    and the whole quadratic form
$ \omega_{\kappa}[R_{-\Lambda}v] $
    tend to zero for
$ \Lambda \to \infty $.

    Summarizing the above we have shown that the quadratic form of the operator
$ L_{\kappa}^{1/2} $
    can be reconstructed by its action on the set
$ \DD(\lL) $.
    The same proof is valid for quadratic forms of square roots of
    positive extenstions of the transverse Laplace operator,
    one just needs to use corresponding antisymmetric potential
    and equation
(\ref{gap2})
    instead of
(\ref{gap1}).

    We would like to note that similar construction does not work
    for extensions of the quadratic form of Laplacian generated by
    the interaction with
$ \delta $-potential.
    In this case the quadratic form corresponding to
(\ref{wkRv})
    and expressions
(\ref{L1}),
(\ref{L2})
    are dimensionless
    and counterpart of the last integral does not tend to zero for
$ \Lambda \to \infty $.
    That is we may suppose that the closure of the quadratic form
$ \omega_{\kappa} $
    acting on the set
$ \DD(\lL) $
    represents a restriction
    of the form
$ \omega_{\kappa} $
\begin{equation*}
    \overline{\omega_{\kappa}|_{\DD(\lL)}} =
    \tilde{\omega}_{\kappa}
	\subset \omega_{\kappa} \quad
    \tilde{\omega}_{\kappa} \neq \omega_{\kappa}.
\end{equation*}
    This restriction corresponds to a self-adjoint operator different from
$ L_{\kappa}^{1/2} $
    such that quadratic form of its square is not an extension of the
    quadratic form of Laplacian. And this means that equation
(\ref{lqeq})
    is not apparently valid for quadratic forms extended by means of the interaction with
$ \delta $-potential.

    Turning to the other potentials of the general form
\begin{equation*}
    \hat{v}_{\alpha}(\vec{p}) = p^{\alpha},
\end{equation*}
    we propose that,
    besides the case
$ \alpha = -1/2 $,
    the statement of this section is valid for
\begin{equation*}
    -\frac{1}{2} < \alpha < 0 ,
\end{equation*}
    as long as expressions
(\ref{L1}),
(\ref{L2})
    are proportional to
$ \Lambda^{\alpha} $
    and thus tend to zero for
$ \Lambda \to \infty $.
    But we are choosing
$ \alpha = -1/2 $
    for our model by the reason that in this case
    vectors
$ R_{\rho} v $
    are the closest functions to the domain
$ \DD(\lL) $.
    And only this case has logarithmic Nevanlinna functions and in this way
    exhibits confinement of singularities in the potential.

\section{Hilbert space and eigenvalues}
\subsection{Normalization of eigenstates}
\label{S:Normalization}
    Further development of constructed Gaussians
$ \Omega_{\kappa} $
    involves introduction of the scalar product and in particular
    investigation of the possibility to normalize the functionals
    with respect to the ground state of the free theory
$ \Omega_{0} $.
    Such a normalization can be performed by means of the functional
    integration, the result is proportional to
\begin{equation*}
    \frac{\langle \Omega_{\kappa} | \Omega_{\kappa} \rangle}{
        \langle \Omega_{0} | \Omega_{0} \rangle}
    = \frac{\int\exp\{-\omega_{\kappa}[A]\}
        \,\prod_{\vec{x}}\delta A(\vec{x})}{\int\exp\{-\omega[A]\}
        \,\prod_{\vec{x}}\delta A(\vec{x})}
    = \exp \{\frac{1}{4} \Tr(\ln L-\ln L_{\kappa}) \}.
\end{equation*}
    Using the equation
\begin{equation*}
    (R_{\bar{\mu}}v, R_{\mu}v) = \frac{\pl}{\pl \mu}(v, R_{\mu}v)
	= \sigmab'(\mu),
\end{equation*}
    derived by differentiation of
(\ref{sigmaDef}),
    it is not hard to see,
    from the representation of
$ \ln L_{\kappa} $
    similar to
(\ref{Lkappa12}),
    that the trace of the difference of logarithms can
    be written as the following expression
\begin{equation}
\label{intTrLn}
    \Tr (\ln L - \ln L_{\kappa}) = -\frac{1}{2\pi i} \int_{0}^{\infty} \bigl(
	\frac{\sigmab'(\lambda +i0)}{\kappa - \sigmab(\lambda +i0)}
	- \frac{\sigmab'(\lambda -i0)}{\kappa - \sigmab(\lambda -i0)}
    \bigr) \ln \lambda \, d\lambda.
\end{equation}
    The latter can be further estimated as the integral
\begin{multline}
    \Tr (\ln L - \ln L_{\kappa}) \sim \\
\label{intTrLnInf}
    \sim \frac{1}{2\pi i} \int_{0}^{\infty}\frac{
	\sigmab(\lambda-i0) - \sigmab(\lambda+i0)}{(
    \kappa-\sigmab(\lambda+i0))(\kappa-\sigmab(\lambda-i0))}
    \sigmab'_{\text{ln}}(\lambda)\ln \lambda \, d\lambda \sim 
    \int_{C_{\text{IR}}}^{\infty}
	\frac{d\lambda}{\lambda \ln\lambda},
\end{multline}
    which diverges as dilogarithm of the upper limit.
    Here
$ \sigmab'_{\text{ln}}(\lambda) $
    is the term of
$ \sigmab'(\lambda) $
    stemming from the logarithm in
$ \sigma(\lambda) $ --
    it is the first term in the decomposition
\begin{multline}
\label{sigmal}
    \frac{\pl\sigmab}{\pl \lambda} = -\frac{2\pi}{\lambda}
	+\frac{2\pi}{\lambda^{3/2}x}
	(\cos\sqrt{\lambda}x \Si\sqrt{\lambda}x
	    - \sin\sqrt{\lambda}x\Ci\sqrt{\lambda}x) +\\
	+ \frac{2\pi}{\lambda}
	(\sin\sqrt{\lambda}x \Si\sqrt{\lambda}x
	    + \cos\sqrt{\lambda}x\Ci\sqrt{\lambda}x)
\end{multline}
    for
$ \sigmab(\lambda) $
    from
(\ref{gmk}) or the first term in the decomposition
\begin{multline}
    \frac{\pl\sigmab^{v,u}_{-}}{\pl\lambda}
	= -\frac{4\pi}{3\lambda} - \frac{12\pi}{\lambda^{2}x^{2}}
 + \bigl(\frac{12\pi}{\lambda^{5/2}x^{3}}-\frac{4\pi}{\lambda^{3/2}x}\bigr)
    \bigl(\cos\sqrt{\lambda}x \Si\sqrt{\lambda}x
            - \sin\sqrt{\lambda}x\Ci\sqrt{\lambda}x\bigr) \\
\label{sigmapl}
    +\frac{12\pi}{\lambda^{2} x^{2}}\bigl(
	\sin\sqrt{\lambda} x \Si\sqrt{\lambda} x
	+\cos\sqrt{\lambda} x \Ci\sqrt{\lambda} x \bigr),
\end{multline}
    for
$ \sigmab^{v,u}_{-}(\lambda) $
    from
(\ref{sigmavum}).
    We can see that contributions to
(\ref{intTrLn})
    from other terms in
(\ref{sigmal}),
(\ref{sigmapl})
    are finite and that the infinity of
(\ref{intTrLnInf})
    is eliminated when subtracting traces of logarithms of
$ L_{\kappa} $
    for sufficiently close pairs
$ \vec{x}_{1} $,
$ \vec{x}_{2} $
    and
$ \vec{x}_{3} $,
$ \vec{x}_{4} $
\begin{equation*}
    |\Tr \ln L_{\kappa;\vec{x}_{1},\vec{x}_{2}} -
    \Tr \ln L_{\kappa;\vec{x}_{3},\vec{x}_{4}} | < \infty .
\end{equation*}
    The infinitesimal version of this statement also holds,
    that is the derivative
\begin{multline}
    \frac{\pl \Tr\ln L_{\kappa;\vec{x}_{1},\vec{x}_{2}}}{\pl x}
    = \frac{1}{2\pi i} \int_{0}^{\infty} \Bigl(
	\bigl(
    \frac{\pl_{x}\pl_{\lambda} \sigmab}{
	\kappa - \sigmab}\bigr|_{\lambda+i0}-
    \frac{\pl_{x}\pl_{\lambda} \sigmab}{
	\kappa - \sigmab}\bigr|_{\lambda-i0}
	\bigr) -\\
\label{plTrLn}
    - \bigl(
\frac{\pl_{x}\sigmab \pl_{\lambda}\sigmab}{
    (\kappa -\sigmab)^{2}}\bigr|_{\lambda+i0}-
\frac{\pl_{x}\sigmab \pl_{\lambda}\sigmab}{
    (\kappa -\sigmab)^{2}}|_{\lambda-i0}
    \bigr) \Bigr)
    \ln \lambda \, d\lambda
\end{multline}
    is finite for
$ x = |\vec{x}_{1}-\vec{x}_{2}| $.
    This finiteness can be directly verified for
$ \sigmab^{v,u}_{-}(\lambda) $,
    but for
$ \sigmab(\lambda) $
    from
(\ref{gmk})
    some integration by parts
    is required in
(\ref{intTrLn})
    before applying derivative \textit{w.\,r.}~to
$ x $.
    Here we assume that
$ \kappa $
    does not depend on 
$ \vec{x}_{1} $,
$ \vec{x}_{2} $
    however this may not be true for the specific physical model.

    As was stated in
(\ref{sigmazero})
    Nevanlinna functions from denominators in
(\ref{intTrLn})
    and
(\ref{plTrLn})
    have a logarithmic branch point multiplied by zero at
$ \lambda = 0 $
    for the distance
$ |\vec{x}_{2}-\vec{x}_{1}| $
    tending to the limit
$ x_{\mathrm{b}} $
    imposed by the absence of negative eigenvalues of
$ L_{\kappa} $.
    This results in an infinite growth of the norm
    of the eigenstate corresponding to
$ \vec{x}_{1} $,
$ \vec{x}_{2} $
    for which
$ |\vec{x}_{2}-\vec{x}_{1}| = x_{\text{b}} $
\begin{equation*}
    \frac{\langle \Omega_{\kappa,x_{\mathrm{b}}} |
	\Omega_{\kappa,x_{\mathrm{b}}} \rangle}{
	\langle \Omega_{\kappa, x} | \Omega_{\kappa, x} \rangle}
    = \exp\{\frac{1}{4}\Tr (\ln L_{\kappa,x}- \ln L_{\kappa,x_{\mathrm{b}}})\}
    \sim \exp\{\frac{1}{4} \int_{0}^{C_{\text{UV}}}
	\frac{d\lambda}{\lambda \ln \lambda}\} .
\end{equation*}
    When applied to
(\ref{intTrLnInf})
    the above equation gives the estimate
\begin{equation*}
    \frac{\langle \Omega_{\kappa,x_{\mathrm{b}}} |
	\Omega_{\kappa,x_{\mathrm{b}}} \rangle}{
	\langle \Omega_{0} | \Omega_{0} \rangle}
    \sim \exp\{\frac{1}{4} \Bigl(
	\int_{0}^{C_{\text{UV}}}
	+ \int_{C_{\text{IR}}}^{\infty}
	\Bigr)\frac{d\lambda}{\lambda \ln \lambda}\} ,
\end{equation*}
    that is we have a sum of divergencies of the opposite signs
    with equal coefficients in the exponent which requires
    additional analysis.

\subsection{Overlap with the ground state of the free field}
    We have seen that Gaussians
$ \Omega_{\kappa} $
    have an infinite norm \textit{w.\,r.}~to
    the norm of the ground state of the free field
$ \Omega_{0} $.
    This however does not allow to decide whether the former and the latter
    belong to the same Hilbert space or not. The question is solved by
    means of the mutual normalized scalar product
    called overlap
    by R.~Jackiw
\cite{Jackiw}
\begin{equation}
\label{overlapi}
    \frac{\langle \Omega_{\kappa} | \Omega_{0} \rangle}{
    \langle \Omega_{\kappa} | \Omega_{\kappa} \rangle^{1/2}
    \langle \Omega_{0} | \Omega_{0} \rangle^{1/2}}
    = \exp\{-\Tr \ln\frac{1}{2}\Bigl(
    \frac{L_{\kappa}^{1/4}}{L^{{1/4}}} + \frac{L^{1/4}}{L_{\kappa}^{1/4}} 
    \Bigr)\}.
\end{equation}
    Zero overlap, that is infinite trace in the RHS, means that
    Gaussian functionals
$ \Omega_{0} $ and
$ \Omega_{\kappa} $
    reside in different Hilbert spaces.

    In order to estimate the overlap let us write the formal representation
\begin{equation*}
    \frac{L_{\kappa}^{1/2}}{L^{{1/2}}} = \II + \HSop, \quad
    \HSop = \frac{L_{\kappa}^{1/2}-L^{1/2}}{L^{1/2}}, \quad
    \HSop^{*} = \HSop
\end{equation*}
    and fix the ordering of non-commuting operators for example by
\begin{equation}
\label{HSop}
    \HSop = L^{-1/4}(L_{\kappa}^{1/2}-L^{1/2}) L^{-1/4}.
\end{equation}
    We have no information about the spectrum of
$ \HSop $,
    we only know that it is real and located to the right
    of point -1, as long as the operator
$ L^{-1/4} L_{\kappa}^{1/2} L^{-1/4} $
    is positive.
    The trace under the logarithm in the RHS of
(\ref{overlapi})
    is clearly a function of
$ \HSop $
    so that we can treat it as a scalar and write
\begin{multline*}
    \Tr \ln\frac{1}{2}\Bigl(
    \frac{L_{\kappa}^{1/4}}{L^{{1/4}}} + \frac{L^{1/4}}{L_{\kappa}^{1/4}} 
    \Bigr)
    = \Tr \ln \frac{1}{2}\bigl((\II+\HSop)^{1/2} + (\II+\HSop)^{-1/2}\bigr) =\\
    = \Tr \ln \bigl(\II + \frac{1}{2}
	\bigl((\II+\HSop)^{1/2} + (\II+\HSop)^{-1/2} -2\II\bigr)\bigr) \leq \\
    \leq \frac{1}{2} \Tr
        \bigl((\II+\HSop)^{1/2} + (\II+\HSop)^{-1/2} -2\II\bigr) .
\end{multline*}
    This inequality stems from the behaviour of the logarithm
    in the vicinity of unity and it holds when the additive to
    the unity
    in the second line is positive, which is equivalent to
\begin{equation}
\label{deltain}
    0 \leq \frac{\HSop^{2}}{\II+\HSop} .
\end{equation}
    Further we may use the algebraic inequality
\begin{equation*}
    \frac{1}{2} \Tr
        \bigl((\II+\HSop)^{1/2} + (\II+\HSop)^{-1/2} -2\II\bigr) 
    \leq \frac{1}{4} \Tr \bigl(\HSop + (\II+\HSop)^{-1} - \II\bigr),
\end{equation*}
    which also requires just
(\ref{deltain})
    to hold. Then we can turn back to operators
$ L^{1/2} $,
$ L_{\kappa}^{1/2} $
    and express
\begin{multline}
\label{TrLL}
    \Tr \bigl(\HSop + (\II+\HSop)^{-1} - \II\bigr)
    = \Tr \bigl(\frac{L_{\kappa}^{1/2}-L^{1/2}}{L^{1/2}}
	+ \frac{L_{\kappa}^{-1/2}}{L^{-1/2}} - \II \bigr) = \\
    = \Tr \bigl(\frac{L_{\kappa}^{1/2}-L^{1/2}}{L^{1/2}}
	+ \frac{L_{\kappa}^{-1/2} - L^{-1/2}}{L_{\kappa}^{-1/2}}\bigr).
\end{multline}
    Note that the ordering
(\ref{HSop})
    does not matter here since there are only two factors under the trace
    in each term.
    Now we have equation to be estimated by means of Dunford integrals.
    The first term in
(\ref{TrLL})
    is the trace of
$ \HSop $
    and it can be calculated by prepending
(\ref{Lkappa12})
    with the factor
$ p^{-1} $
\begin{multline}
\label{TrLp}
    \Tr \frac{L_{\kappa}^{1/2}-L^{1/2}}{L^{1/2}} = \Tr \HSop = \\
    = \frac{1}{2\pi i} \Tr \int_{0}^{\infty} \bigl(
	\frac{\frac{1}{p} R_{\lambda+i0}v(R_{\lambda-i0}v, \, \cdot\, )}{
	\kappa-\sigmab(\lambda+i0)}
	-\frac{\frac{1}{p} R_{\lambda-i0}v(R_{\lambda+i0}v, \, \cdot\, )}{
	\kappa-\sigmab(\lambda-i0)}
	\bigr) \lambda^{1/2} \,d\lambda = \\
    = \frac{1}{2\pi i} \int_{0}^{\infty} \bigl(
	\frac{(R_{\lambda-i0}v, \frac{1}{p} R_{\lambda+i0}v)}{
	\kappa-\sigmab(\lambda+i0)}
	-\frac{(R_{\lambda+i0}v, \frac{1}{p} R_{\lambda-i0}v)}{
	\kappa-\sigmab(\lambda-i0)}
	\bigr) \lambda^{1/2} \,d\lambda .
\end{multline}
    The second term in
(\ref{TrLL})
    can be processed in a similar way
\begin{multline}
\label{TrLm}
    \Tr \frac{L_{\kappa}^{-1/2}-L^{-1/2}}{L^{-1/2}} = \\
    = \frac{1}{2\pi i} \int_{0}^{\infty} \bigl(
	\frac{(R_{\lambda-i0}v, p R_{\lambda+i0}v)}{
	\kappa-\sigmab(\lambda+i0)}
	-\frac{(R_{\lambda+i0}v, p R_{\lambda-i0}v)}{
	\kappa-\sigmab(\lambda-i0)}
	\bigr) \lambda^{-1/2} \,d\lambda .
\end{multline}
    Straightforward calculations show that for the case of the scalar field
\begin{align}
\label{RRp}
    (R_{\bar{\mu}}v, p R_{\mu}v) &= 2\pi^{2}\bigl(\frac{1}{2\sqrt{-\mu}}
	- \frac{e^{-\sqrt{-\mu}x}}{2\sqrt{-\mu}}\bigr), \\
\label{RRm}
    (R_{\bar{\mu}}v, \frac{1}{p} R_{\mu}v) &= 2\pi^{2}\bigl(\frac{1}{2(\sqrt{-\mu})^{3}}
	- \frac{1}{\mu^{2}x} + \frac{e^{-\sqrt{-\mu}x}}{\mu^{2}x}
	- \frac{e^{-\sqrt{-\mu}x}}{2(\sqrt{-\mu})^{3}} \bigr),
\end{align}
    where
$ \sqrt{\mu} $
    has cut along the negative semi-axis and is positive for positive
$ \mu $,
    so that
\begin{equation*}
    \sqrt{-(\lambda\pm i0)} = \mp i \lambda^{1/2}, \quad \lambda \geq 0.
\end{equation*}
    We can see that the first terms in
(\ref{RRp}),
(\ref{RRm})
    make integrals
(\ref{TrLp}),
(\ref{TrLm})
    diverge in the infinity as dilogarithm of the upper bound. This means that
$ \HSop $,
    similarly to
$ \ln(\II+\HSop) $ from
(\ref{intTrLn}),
    is not a trace-class operator, but rather a Hilbert-Schmidt one,
    see
f.\,e.~\cite{GGK}.
    Indeed, the sum of
(\ref{TrLp})
    and
(\ref{TrLm})
    starts from
$ \Tr \HSop^{2} $
    and it is finite
\begin{multline*}
    \Tr \bigl(\frac{L_{\kappa}^{1/2}-L^{1/2}}{L^{1/2}}
	+ \frac{L_{\kappa}^{-1/2} - L^{-1/2}}{L_{\kappa}^{-1/2}}\bigr) =\\
    = \IM\, \frac{\pi}{2} \int_{0}^{\infty} (\kappa-\sigmab(\lambda+i0))^{-1}
	\bigl(\frac{e^{i\sqrt{\lambda}x}-1}{\lambda^{3/2}x} - \frac{ie^{i\sqrt{\lambda}x}}{\lambda} \bigr)
    d\lambda < \infty .
\end{multline*}
    This integral is finite since its first term converges absolutely in
    the infinity while the second one is oscillating fast enough to converge
    conditionally.
    Divergency in the vicinity of 0 is also absent, provided that the
    distance
$ x $
    does not approach the boundary value
$ x_{\text{b}} $.
    Thus we may conclude that overlap
(\ref{overlapi})
    is finite and Gaussians
$ \Omega_{0} $,
$ \Omega_{\kappa} $
    reside in the same Hilbert space.
    
\subsection{Comparison of eigenstates and eigenvalues}
    Calculations presented in the previous section show that
    eigenstates
$ \Omega_{\kappa}(A) $
    and the Gaussian
$ \Omega_{0} $
    reside in the common functional Hilbert space
$ \Hilbert $.
    This space is
    generated by excitations of
$ \Omega_{0} $,
    but can also be represented as Hilbert space with basis of
    excitations of Gaussian
$ \Omega_{\kappa}^{\vec{x}_{1},\vec{x}_{2}}(A) $,
    its elements have the form
\begin{equation}
\label{OmegaExp}
    \Omega(A) = \sum_{n, \vec{p}_{1},\ldots\vec{p}_{n}}
	\psi(\vec{p}_{1},\ldots\vec{p}_{n})
	\Psi_{\kappa}^{\dag}(\vec{p}_{1})\dots\Psi_{\kappa}^{\dag}(\vec{p}_{n})
    \, \Omega_{\kappa}^{\vec{x}_{1},\vec{x}_{2}}(A),
\end{equation}
    where
$ \Psi_{\kappa}^{\dag}(\vec{p}) $
    are creation operators in the sense of Friedrichs
\cite{Friedrichs}
    depending on
$ \vec{x}_{1} $,
$ \vec{x}_{2} $
    and
$ \kappa $
\begin{equation*}
    \Psi_{\kappa;\vec{x}_{1},\vec{x}_{2}}^{\dag} (\vec{p})
	= L_{\kappa;\vec{x}_{1},\vec{x}_{2}}^{1/4} \hat{A}(\vec{p})
	- L_{\kappa;\vec{x}_{1},\vec{x}_{2}}^{-1/4}
	    \frac{\delta}{\delta \hat{A}(\vec{p})}.
\end{equation*}

    At the same time functionals
$ \Omega_{\kappa}^{\vec{x}_{1},\vec{x}_{2}} $ 
    and
$ \Omega_{\kappa}^{\vec{x}_{3},\vec{x}_{4}} $
    have different behaviour in the vicinity of functions
    with expansion
(\ref{ABound})
    in non-coinciding pairs of points
$ \vec{x}_{1} $,
$ \vec{x}_{2} $
    and
$ \vec{x}_{3} $,
$ \vec{x}_{4} $.
    Hence they are eigenfunctionals of different self-adjoint
    operators in the same space
$ \Hilbert $,
    denote them as
$ \HH_{\kappa}^{\vec{x}_{1},\vec{x}_{2}} $
    and
$ \HH_{\kappa}^{\vec{x}_{1},\vec{x}_{2}} $.
    These operators
    are defined on different domains,
    so that the action of
$ \HH_{\kappa}^{\vec{x}_{1},\vec{x}_{2}} $
    is not defined on functional
$ \Omega_{\kappa}^{\vec{x}_{3},\vec{x}_{4}} $
    and vice versa.
    But it turns out that it is
    possible to construct a model which includes as dynamical variables
    not only the quantum field
$ \hat{A}(\vec{p}) $
    but also centers of external sources
$ \vec{x}_{1} $ and
$ \vec{x}_{2} $.
    For this one needs to define action of
$ \HH_{\kappa}^{\vec{x}_{1},\vec{x}_{2}} $
    at least on functionals generated by excitations of
$ \Omega_{\kappa}^{\vec{x}_{3},\vec{x}_{4}} $
    for sufficiently close pairs
$ \vec{x}_{1} $,
$ \vec{x}_{2} $
    and
$ \vec{x}_{3} $,
$ \vec{x}_{4} $.

    We cannot identify the action
$ \HH_{\kappa}^{\vec{x}_{1},\vec{x}_{2}}
    \Omega_{\kappa}^{\vec{x}_{3},\vec{x}_{4}} $
    as an element of
$ \Hilbert $,
    because the appearance of ``additional''
    eigenfunctional contradicts the completeness
    of the set of excitations of 
$ \Omega_{\kappa}^{\vec{x}_{1},\vec{x}_{2}} $
    from
(\ref{OmegaExp}).
    But we can decompose
$ \Omega_{\kappa}^{\vec{x}_{3},\vec{x}_{4}} $
    via ecxitations of
$ \Omega_{\kappa}^{\vec{x}_{1},\vec{x}_{2}} $
    and calculate the action of
$ \HH_{\kappa}^{\vec{x}_{1},\vec{x}_{2}} $
    in the term-by-term way
\begin{equation}
\label{HOmega}
    \HH_{\kappa}^{\vec{x}_{1},\vec{x}_{2}}
    \Omega_{\kappa}^{\vec{x}_{3},\vec{x}_{4}}
    = \sum_{n,\vec{p}_{1},\ldots\vec{p}_{n}}\psi(\vec{p}_{1},\ldots\vec{p}_{n})
    \HH_{\kappa}^{\vec{x}_{1},\vec{x}_{2}}
	\Psi_{\kappa}^{\dag}(\vec{p}_{1})\dots\Psi_{\kappa}^{\dag}(\vec{p}_{n})
    \, \Omega_{\kappa}^{\vec{x}_{1},\vec{x}_{2}}(A),
\end{equation}
    and then use this equation to define expectation (quadratic form)
\begin{equation}
\label{Osp}
    \langle \Omega_{\kappa}^{\vec{x}_{3},\vec{x}_{4}} | 
    \HH_{\kappa}^{\vec{x}_{1},\vec{x}_{2}}
    | \Omega_{\kappa}^{\vec{x}_{3},\vec{x}_{4}} \rangle ,
\end{equation}
    which has to be finite.
    To justify this assumption one can remind the finite-dimensional theory
    of singular perturbations of differential operators
\cite{AK}.
    In this subject different perturbations of the same operator
    are defined on different domains but quadratic forms like
(\ref{Osp})
    are finite on the set which includes domains of all the perturbations.
    That is we can fix some functional
$ \Omega(A) $
    and then using different expansions
\begin{align*}
    \Omega(A) &= \sum_{n, \vec{p}_{1},\ldots\vec{p}_{n}}
	\psi_{\vec{x}_{1},\vec{x}_{2}}(\vec{p}_{1},\ldots\vec{p}_{n})
	\Psi_{\kappa;\vec{x}_{1},\vec{x}_{2}}^{\dag}(\vec{p}_{1})
	    \dots\Psi_{\kappa;\vec{x}_{1},\vec{x}_{2}}^{\dag}(\vec{p}_{n})
    \, \Omega_{\kappa}^{\vec{x}_{1},\vec{x}_{2}}(A) =\\
    &= \sum_{n, \vec{p}_{1},\ldots\vec{p}_{n}}
	\psi_{\vec{x}_{3},\vec{x}_{4}}(\vec{p}_{1},\ldots\vec{p}_{n})
	\Psi_{\kappa;\vec{x}_{3},\vec{x}_{4}}^{\dag}(\vec{p}_{1})
	    \dots\Psi_{\kappa;\vec{x}_{3},\vec{x}_{4}}^{\dag}(\vec{p}_{n})
    \, \Omega_{\kappa}^{\vec{x}_{3},\vec{x}_{4}}(A)
\end{align*}
    define on it actions of operators
$ \HH_{\kappa}^{\vec{x}_{1},\vec{x}_{2}} $
    with different positions of
$ \vec{x}_{1} $,
$ \vec{x}_{2} $
    and after that calculate expectation values
$    \langle \Omega | 
    \HH_{\kappa}^{\vec{x}_{1},\vec{x}_{2}}
    | \Omega \rangle 
$.
    Important condition for such calculations is the possibility to
    uniformly define eigenvalues
$ \Tr L_{\kappa;\vec{x}_{1},\vec{x}_{2}}^{1/2} $
    of functionals
$ \Omega_{\kappa}^{\vec{x}_{1},\vec{x}_{2}} $
    for different pairs
$ \vec{x}_{1} $,
$ \vec{x}_{2} $.
    Dunford integral for the difference of traces
$ \Tr L_{\kappa}^{1/2} - \Tr L^{1/2} $
    diverges even worse than
(\ref{intTrLn})
\begin{multline}
    \Tr (L_{\kappa}^{1/2} - L^{1/2})
    = \frac{1}{2\pi i} \int_{0}^{\infty} \bigl(
	\frac{\sigmab'(\lambda +i0)}{\kappa - \sigmab(\lambda +i0)}
	- \frac{\sigmab'(\lambda -i0)}{\kappa - \sigmab(\lambda -i0)}
    \bigr) \lambda^{1/2} \, d\lambda \sim \\
\label{plTr12Div}
    \sim \frac{1}{2\pi i} \int_{0}^{\infty}\frac{
	(\sigmab(\lambda+i0) - \sigmab(\lambda-i0))
    \sigmab'_{\text{ln}}(\lambda) \lambda^{1/2}}{(
    \kappa-\sigmab(\lambda+i0))(\kappa-\sigmab(\lambda-i0))}
    \, d\lambda \sim 
    - \int_{C_{\text{IR}}}^{\infty}
	\frac{d\lambda}{\lambda^{1/2} \ln^{2}\lambda}.
\end{multline}
    However the derivative of this integral \textit{w.\,r.}~to
    the distance between
$ \vec{x}_{1} $
    and
$ \vec{x}_{2} $
\begin{multline}
    \frac{\pl \Tr L_{\kappa;\vec{x}_{1},\vec{x}_{2}}^{1/2}}{\pl x} 
    = \frac{1}{2\pi i} \int_{0}^{\infty} \Bigl(
	\bigl(
\frac{\pl_{x}\pl_{\lambda} \sigmab}{\kappa - \sigmab}(\lambda+i0)-
\frac{\pl_{x}\pl_{\lambda} \sigmab}{\kappa - \sigmab}(\lambda-i0)
    \bigr) -\\
\label{plTr12}
    - \bigl(
\frac{\pl_{x}\sigmab \pl_{\lambda}\sigmab}{
	(\kappa -\sigmab)^{2}}(\lambda+i0)-
\frac{\pl_{x}\sigmab \pl_{\lambda}\sigmab}{
	(\kappa -\sigmab)^{2}}(\lambda-i0)
    \bigr) \Bigr)
    \lambda^{1/2} \, d\lambda
\end{multline}
    as well as the difference
\begin{equation*}
    \Tr L_{\kappa;\vec{x}_{1},\vec{x}_{2}}^{1/2} -
    \Tr  L_{\kappa;\vec{x}_{3},\vec{x}_{4}}^{1/2}  
\end{equation*}
    is finite.
    Here consideration provided for traces of logarithms in the Section
\ref{S:Normalization}
    also works.
    That is only the first terms
    from
(\ref{sigmal})
    and
(\ref{sigmapl})
    bring divergency to
(\ref{plTr12Div}),
    and the latter is generated by the constant in the gap
(\ref{gap1}) or
(\ref{gap2}) 
    and in this way does not depend on
$ \vec{x}_{1} $ and
$ \vec{x}_{2} $.

    The noticeable property of divergency
(\ref{plTr12Div})
    is that it is negative and hence
    it creates an infinitely deep energy well,
    which separates functionals of the presented model from excitations
    of the ground state of the free field.
    And infinitely deeper wells are created when extending quadratic form
    of Laplacian with more singular vectors in
(\ref{Dlk}),
    for example by considering model with several pairs or triples of
    external sources.

\section{Summary}    
    We have shown that in order to construct alternative
    solutions to the eigenstate
    equation of the free quantum theory the external potential
    incorporating at least two connected singularities has to be used.
    The distance between the latter
    is proved to be limited by the extension
    parameter.
    We have shown for the special type of singularities
    that the quadratic
    form of the square root of perturbation of Laplacian 
    is reconstructed by its action on the domain
    of the unperturbed operator.
    We have estimated spectral integrals for traces of logarithms
    and argue that constructed eigenfunctionals belong to
    the same Hilbert space
    as excitations of the free ground state but have infinitely lower
    energy.
    We have also shown that corresponding eigenvalues of quantum
    Hamiltonians depend continuously
    on locations of the centers of singularities.


\begin{thebibliography}{00}

\bibitem{Hatfield}
\printbookitem{B.~Hatfield}{1992}
    {Free Fields in the Schrodinger Representation. In:
    Quantum Field Theory of Point Particles and Strings}
    {Reading, Massachusetts: Addison Wesley Longman}{199--224}

\bibitem{Koshmanenko1}
\printbookitem{V.~Koshmanenko}{1999}
{Singular Quadratic Forms in Perturbation Theory}
{Dordrecht: Springer Science+Business Media}{59--226}

\bibitem{TB}
\printpreprintitem{T.~A.~Bolokhov}{2019}
{Singular perturbations of a free quantum field Hamiltonian}
{arXiv:1912.01458 [math-ph]}

\bibitem{Jackiw}
\printpreprintitem{R.~Jackiw}{1987}
{Functional Representation for Quantized Fields}
{MIT-CTP-1511; Contribution to: 1st Asia Pacific Conference on High-energy Physics: Superstrings, Anomalies and Field Theory}

\bibitem{BKV}
\printjournalitem{A.~Alonso and B.~Simon}{1980}
{The Birman - Krein - Vishik theory of selfadjoint
    extensions of semibounded operators}
{J. Operator Theory}{4}{251--270}

\bibitem{KS}
\printjournalitem{A.~Kiselev and B.~Simon}{1995}
{Rank one perturbations with infinitesimal coupling}
{J. Funct. Anal.}{130}{345--356}

\bibitem{AK}
\printbookitem{S.~Albeverio and P.~Kurasov}{2000}
{Singular Perturbation of Differential Operators.
    Solvable Schr\"odinger type Operators}
{Cambridge: Cambridge University Press}{9--62}

\bibitem{Krein}
\printjournalitem[(62)]{M.~G.~Krein}{1947}
{The theory of self-adjoint extensions of semi-bounded
    Hermitian transformations and its applications}
{Rec. Math. (Mat. Sbornik) N.S.}{20}{431--495}

\bibitem{GT}
\printpreprintitem{F.~Gesztesy and E.~Tsekanovskii}{1997}
{On Matrix-Valued Herglotz Functions}
{arXiv:funct-an/9712004}

\bibitem{LFres}
\printjournalitem[1]{L.~D.~Faddeev}{2006}
{Notes on divergences and dimensional transmutation in Yang-Mills theory}
{Theor.\ Math.\ Phys.}{148}{986--994}

\bibitem{TBZ}
\printjournalitem{T.~A.~Bolokhov}{2024}
{Correlation functions of two 3-dimensional transverse potentials with power singularities}
{Zap.\ Nauch. Sem. POMI}{532}{109--118}

\bibitem{Riesz}
\printjournalitem{F.~Riesz and E.~R.~Lorch}{1936}
{The integral representation of unbounded self-adjoint transformations in Hilbert space}
{Trans Amer. Math. Soc.}{39}{331-340}

\bibitem{DS}
\printbookitem{N.~Dunford and J.~T.~Schwartz}{1958}
{General Spectral Theory. In: Linear Operators 1}
{New York: Interscience Publishers}{555--612}

\bibitem{GGK}
\printbookitem{I.~Gohberg, S.~Goldberg and N.~Krupnik}{2000}
{Trace Class and Hilbert-Schmidt Operators in Hilbert Space. In: Traces and Determinants of Linear Operators}
{Operator Theory Advances and Applications, vol 116. Birkhuser, Basel}{47--90}

\bibitem{Friedrichs}
\printjournalitem{K.~O.~Friedrichs}{1951}
{Mathematical aspects of the quantum theory of fields I-II}
{Comm. Pure Appl. Math.}{4}{161--224}

\end{thebibliography}
\end{document}